\pdfoutput=1
\documentclass[a4paper,10pt]{article}
\usepackage{fullpage}
\usepackage{hyperref}
\usepackage{graphicx,subfigure}
\usepackage{amsmath,amssymb,dsfont}

\usepackage{algorithmic,algorithm}

\floatname{algorithm}{Procedure}

\algsetup{indent=1em}

\usepackage{listings}
\newcommand{\eg}{\emph{e.g.\/}}
\newcommand{\ie}{\emph{i.e.\/}}

\newcommand{\Mathematica}{\emph{Mathematica}}
\newcommand{\MathLink}{\emph{MathLink}}

\newcommand{\M}{\ensuremath{\mathbb{M}}}
\DeclareMathOperator{\tr}{tr}

\newcommand{\C}{\ensuremath{\mathds{C}}}
\newcommand{\Cplx}{\ensuremath{\C}}

\newcommand{\ket}[1]{\ensuremath{|#1\rangle}}
\newcommand{\bra}[1]{\ensuremath{\langle#1|}}
\newcommand{\ketbra}[2]{\ensuremath{\ket{#1}\bra{#2}}}

\newcommand{\scalar}[2]{\left( #1 , #2 \right)}

\usepackage{xcolor}

\newcommand{\TRQS}{\InlineCode{TRQS}}
\newcommand{\InlineCode}[1]{{\small \textbf{#1}}}

\newcommand{\libquantis}{\texttt{libQuantis}}
\newcommand{\libquantisnoWH}{\texttt{libQuantis-NoHW}}
\newcommand{\NormalDist}[2]{\ensuremath{\mathcal{N}(x,y)}}

\begin{document}

\title{Generating and using truly random quantum states in \Mathematica}

\author{Jaros{\l}aw Adam Miszczak\\Institute of Theoretical and Applied
Informatics,\\ Polish Academy of Sciences, Ba{\l}tycka 5, 44-100 Gliwice,
Poland}
\date{30/07/2011 (v. 0.23)}

\maketitle

\begin{abstract}
The problem of generating random quantum states is of a great interest from the
quantum information theory point of view. In this paper we present a package for
\Mathematica\ computing system harnessing a specific piece of hardware, namely
Quantis quantum random number generator (QRNG), for investigating statistical
properties of quantum states. The described package implements a number of
functions for generating random states, which use Quantis QRNG as a source of
randomness. It also provides procedures which can be used in simulations not
related directly to quantum information processing.



\end{abstract}

\section*{Program Summary}
\noindent\emph{Program title:} TRQS\\
\emph{Program author:} Jaros{\l}aw A. Miszczak\\
\emph{Distribution format:} tar.gz\\
\emph{No. of bytes in distributed program, including test data, etc.:} $8.9 \cdot 10^4$\\
\emph{No. of lines in distributed program, including test data, etc.:} $\sim 1000$\\
\emph{Programming language:} Mathematica, C\\
\emph{Computer:} any supporting the recent version of Mathematica\\
\emph{Operating system:} any platform supporting Mathematica; tested with GNU/Linux (32 and 64 bit)\\
\emph{RAM:} case-dependent\\
\emph{Nature of problem:} Generation of random density matrices.\\
\emph{Method of solution:} Use of a physical quantum random number generator.

\section{Introduction}
As full scale quantum computing devices are still missing, the simulation of
quantum computers has gained considerable attention as a method for
investigation the behaviour of quantum algorithms and
protocols~\cite{quantiki-simulators}. It also provides a valuable method for
inspecting the mathematical structure of quantum theory by providing information
about statistical properties of quantum states and operations~\cite{BZ06}.

Generating random numbers using the statistical nature of quantum theory
provides one of the first practical applications of quantum information theory.
At the same time the high quality of random numbers generated using quantum
random number generators is based on the very basic principles of
Nature~\cite{calude10experimental}. Since random numbers are important in many
areas of human activity, at the moment this provides one of the most important
applications of quantum information theory.

During the last several years many simulators of quantum computers have been
developed and the most up-to-date list of available software is available
at~\cite{quantiki-simulators}. Many simulators of quantum information processing
were developed using \Mathematica\ computing system~\cite{qucalc, qdensity,
quantum2, QI, qcwave}. Also some attention has been devoted to utilising CUDA
programming model~\cite{gutierrez10cuda} and parallel processing
model~\cite{glendinning03parallelization, deraedt07massively, tabakin09qcmpi}.
The research effort in using parallel and distributed computing for the purpose
of quantum information processing was motivated by the amount of computational
resources needed in order to perform a simulation of quantum computation.

In this paper we present a \TRQS\ (True Quantum Random States) package for
\Mathematica\ computing system harnessing a specific piece of hardware, namely 
Quantis quantum random number generator (QRNG), for investigating statistical
properties of quantum states. The described package implements a number of
functions for using numbers and generating random states (\ie\ random state
vectors and random density matrices), using Quantis QRNG as the source of
randomness. The presented package has been developed for the purpose of quantum
information theory, but it can be easily utilised in other areas of science. 

The motivation for utilising random numbers generated using quantum devices is
twofold. Firstly, QRNGs provide a high-quality source of randomness which can be
used in various areas of computational physics. Recent progress in this
area~\cite{furst10high-speed} suggests that QRNGs are of a great interest for
experimentalists, as well as theorists, working in the field of quantum
information theory. Secondly, it has been shown that the statistical properties
of obtained numbers cannot be reproduced using standard methods of generating
random numbers~\cite{calude10experimental}.

This paper is organised as follows.
In Section~\ref{sec:random-states} we introduce basic theoretical facts
concerning random states and operations.
In Section~\ref{sec:package} we describe the functions implemented in the
presented package and in Section~\ref{sec:examples} we use the described package
to analyse some problems related to quantum information theory. We also use
\TRQS\ package be benchmark the speed of Quantis random number generator. 
Finally, in Section~\ref{sec:conclusions}, we provide some concluding remarks and
discuss the alternative sources of random numbers generated using quantum random
number generators.

\section{Random quantum states}\label{sec:random-states}
The problem of generating random quantum states is of a great interest from the
quantum information theory point of view. Random states appear naturally in many
situations in quantum information processing, especially when one must deal with
the unavoidable interaction of the system in question with the environment.

We start by recalling some basic facts used in the rest of this paper. For a
more complete introduction to mathematical concepts used in quantum information
theory see \eg~\cite{ziman,BZ06}. Next, we present the selected methods of
generating random density matrices implemented in the \TRQS\ package. More
detailed description of the methods for generating random quantum density
matrices can be found in~\cite{zyczkowski10generating}.

\subsection{Basic definitions}
In what follows we restrict our attention to finite-dimensional spaces. We
denote by $\ket{\phi}\in \Cplx^n$ \emph{pure states} \ie\ normalised elements of
the vector space $\Cplx^n$. By $\M_{m,n}$ we denote the set of all $m\times n$
matrices over $\Cplx$ and the set of square $n\times n$ matrices is denoted by
$\M_n$. The set of $n$-dimensional density matrices (normalised, positive
semi-definite operators on~$\Cplx^n$) is denoted by~$\Omega_n$. The set $\M_{n}$
has the structure of a Hilbert space with the scalar product given by
$\scalar{A}{B} = \tr A^\dagger B$. This particular Hilbert space is known as the
Hilbert-Schmidt space of operators acting on $\Cplx^n$ and we will denote it by
$\mathcal{H}_\mathrm{HS}$.

In particular, $\Omega_n\subset\M_{n}$. Moreover, any element of $\Omega_n$ can
be represented as a convex combination (mixture) of one-dimensional projectors.
For any $\rho\in\Omega_n$ there exists a sequence of non-negative numbers
$p_1,p_2,\ldots,p_n$ such that $ \rho=\sum_{i=1}^n p_iP_i, $ where $\{P_i\},\
i=1,2,\ldots,n$ is a sequence of orthonormal one-dimensional projectors. Extreme
elements of the set $\Omega_n$ are exactly the \emph{pure states} and can be
identified with ket vectors $\ket{\psi}\simeq\ketbra{\psi}{\psi}$. Convex
combinations of pure states are refereed to as \emph{mixed states}.

\subsection{Random pure states}
In most cases to describe quantum algorithms and protocols one assumes that it
is possible to avoid unnecessary interactions with the environment~\cite{BZ06}.
In such situation the state of the system remains pure during the evolution,
which is represented by a unitary matrix.

In the case of a pure state (state vectors) there exists a natural measure in
the set, namely the measure generated by the Haar measure on the group of
unitary matrices $\mathrm{U}(n)$. The algorithm for generating random pure
states is presented in Procedure~\ref{alg:pure}. The function
\InlineCode{RandomSimplex(n)} used in this procedure returns an element of a
standard simplex of dimension $n$.

\begin{algorithm}[ht]
\caption{Generation of a random pure state}
\label{alg:pure}
\begin{algorithmic}
\REQUIRE $n\geq 0$
\ENSURE Random pure state $v$ of dimension $n$
\STATE $s \gets$ \InlineCode{RandomSimplex(n)}
\STATE $a[1] \gets \sqrt{s[1]}$
\STATE $p[1] \gets 1$
\STATE $v[1] \gets a[1]*p[1]$
\FOR{$k = 2$ \TO n} 
 \STATE $a[k] \gets \sqrt{s[i]}$
 \STATE $p[k] \gets$ $\exp(i$ \InlineCode{RandomReal}(0,2$\pi$))
    \STATE $v[k] \gets a[k]*p[k]$
\ENDFOR
\RETURN $v$
\end{algorithmic}
\end{algorithm}

Alternatively a random pure states can be obtained by generating a random
unitary matrix and choosing its columns as random pure states.

\subsection{Random mixed states}
The need for using a more general formalism to describe the evolution of quantum
systems is motivated by the fact that in a real-world situation it is impossible
to avoid the interaction of the system with the environment. In this case one
needs to represent the system using quantum channels and introduce density
matrices to describe the state of the system~\cite{BZ06}.

The set of density matrices presents us with more complicated structure than in
the case of pure states. In particular, it is not possible to distinguish one
preferred probability measure in this set and any metric on the set can be used
to introduce one.

The package presented in this paper implements functions for generating random
density matrices distributed according to the probability measure generated by
the Hilbert-Schmidt metric
\begin{equation}\label{eqn:hs-metric}
\|\rho_1-\rho_2\|_{HS} = \sqrt{\tr\left[(\rho_1-\rho_2)^2\right]},
\end{equation}
and 
the Bures metric
\begin{equation}\label{eqn:bures-metric}
\|\rho_1-\rho_2\|_{B} = \sqrt{2-2\sqrt{F(\rho_1,\rho_2)}}
\end{equation}
where $F(\rho_1,\rho_2)$ is a quantum fidelity 
$
F(\rho_1,\rho_2) = \tr|\sqrt{\rho_1}\sqrt{\rho_2}|
$
between two density matrices. In a particular case, when one of the states is
pure, $\rho_1=\ketbra{\psi}{\psi}$, we have
$
F(\ketbra{\phi}{\phi},\rho_2) = \bra{\phi}\rho_2\ket{\psi}
$
and in this case the probability measure is reduced to the Fubini-Study measure.

We also provide a function for generating density matrices distributed with a
family of induced measures~\cite{BZ06}, which can be derived by averaging over
an external subsystem. One should note that the Hilbert-Schmidt measure can be
obtained as an induced measure.

In each case, as a starting point of the algorithm, one needs to use a Ginibre
matrix, \ie\ a complex matrix with elements having real and complex parts
distributed with the normal distribution
\NormalDist{0,1}~\cite{BZ06}. 
\begin{algorithm}
\caption{Generation of the random matrix from the Ginibre ensemble.}
\label{alg:ginibre-matrix}
\begin{algorithmic}
\REQUIRE $m,n>0$
\ENSURE Matrix $G$ of size $m\times n$ 
\FOR{$k=1$ \TO $m$}
    \FOR{$l=1$ \TO $n$}
        \STATE $G[k,l] \gets $ \InlineCode{RandomReal}(0,1) + $i$ \InlineCode{RandomReal}(0,1)
    \ENDFOR
\ENDFOR
\RETURN $G$
\end{algorithmic}
\end{algorithm}

Using \Mathematica\ language, Procedure~\ref{alg:ginibre-matrix} can be written
in a compact form as
\begin{lstlisting}
dist = NormalDistribution[0,1];
GinibreMatrix[m_,n_]:= 
    RandomReal[dist,{m,n}] 
    + I RandomReal[dist,{m,n}]
\end{lstlisting}

\subsubsection{Induced measures and the Hilbert-Schmidt ensemble}\label{sec:induced-measures}
The Hilbert-Schmidt metric defined in Eq.~\ref{eqn:hs-metric} is commonly used
to describe the metric structure of the set of quantum states. This distance
introduces a Euclidean geometry in the space of density matrices. In the special
case of one-qubit density matrices, the space has the form of the Bloch ball.

The Hilbert-Schmidt measure belongs to the class of \emph{induced
measures}~\cite[Ch.14]{BZ06}. In a general case, one can seek a source of
randomness in a given system, by studying the interaction of the $n$-dimensional
system in question with the environment. In such situation the random states to
model the behaviour of the system should be generated by reducing a pure state
in $N\times K$-dimensional space. In what follows we denote the resulting
probability measure by $\mu_{N,K}$.

In particular, the Hilbert-Schmidt probability measure on $n$-dimensional space
$\Omega_n$ can be obtained by reducing a bi-partite pure quantum state from
$\Cplx^{N\times N}$. However, it is easy to see that this measure, as well as
any measure $\mu_{N,K}$, can be obtained by using a simpler procedure.

We start by observing that any complex matrix $X\in\M(\Cplx)$ can be used to
construct a normalised, positive matrix $ \frac{XX^\dagger}{\tr XX^\dagger}$.
Let us now assume that we have a pure state in the $N\times K$-dimensional
Hilbert space, $\ket{\psi}\in\mathcal{H}_N\otimes\mathcal{H}_K=\Cplx^{N\times
K}$. Any such state can be represented in a product basis $ \ket{\psi} =
\sum_{i=1}^N\sum_{j=1}^K X_{ij} \ket{i}\otimes\ket{j},$ where $X\in\M_{N,K}$.
The matrix $XX^\dagger$ is, in this case, equivalent to the partial trace of
$\ketbra{\psi}{\psi}$ with respect to the second subsystem, $
\tr_{\mathcal{H}_K} \ketbra{\psi}{\psi}=XX^\dagger$. Symmetrically we have $
\tr_{\mathcal{H}_N} \ketbra{\psi}{\psi}=X^\dagger X$.

It follows from the above considerations that the spectrum of a density matrix 
obtained using this method is the set of squared and normalised singular
values of the matrix $X$. In particular, if one assumes that the pure states on
$\mathcal{H}_K\otimes\mathcal{H}_N$ are distributed according to the
Fubini-Study measure, the resulting density matrices are distributed with
induced measures. 

In the case of induced measures the elements of the coefficient matrix are
independent random variables and form a Ginibre matrix.

\begin{algorithm}
\caption{Generation of a random density matrix distributed according to induced
probability measure $\mu_{n,k}$ obtained by tracing out the ancillary system of
dimension $k$.}
\label{alg:mixed-induced}
\begin{algorithmic}
\REQUIRE $n \geq 0, k \geq 2$
\ENSURE Random mixed state $\rho$ of dimension $n$
\STATE $G\gets $ \InlineCode{GinibreMatrix(n,k)}
\STATE $\rho\gets GG^\dagger$
\STATE $\rho\gets \frac{1}{\tr\rho}\rho$
\RETURN $\rho$
\end{algorithmic}
\end{algorithm}

In the special case of $K=N$ we obtain the Hilbert-Schmidt ensemble. 


The statistical properties of the set of quantum states with respect to the
probability measure introduced by the Hilbert-Schmidt metric were studied
in~\cite{zyczkowski03hsvolume,zyczkowski05averagefidelity}.


\subsubsection{Bures ensemble}
Another popular measure of the distance between quantum states is the Bures
distance. Its usage is motivated by the fact that this distance, when restricted
to diagonal matrices, is equivalent to the Hellinger distance in statistics,
defined for two discrete probability distributions as
$H(\mathbf{p},\mathbf{q})=\sum_{i}\sqrt{p_iq_i}$. Moreover, the Bures distance
and quantum fidelity are related to the distinguishability of quantum states,
defined as a trace distance between the given states.

The above features distinguish the Bures measure as an optimal method for
generating random density matrices in the situation when no information about
the source of state is present.

The algorithm for generating random density matrices distributed according to
the probability measure based on the Bures distance was provided
in~\cite{osipov10bures}. This algorithm is presented in
Procedure~\ref{alg:mixed-bures}. 

\begin{algorithm}
\caption{Generation of a random density matrix distributed according to
the probability measure induced by the Bures metric}
\label{alg:mixed-bures}
\begin{algorithmic}
\REQUIRE $n \geq 0$
\ENSURE Random mixed state $\rho$ of dimension $n$
\STATE $G\gets $ \InlineCode{GinibreMatrix(n,n)}
\STATE $U\gets $ \InlineCode{RandomUnitary(n)}
\STATE $\rho\gets  (1+U)GG^\dagger(1+U^\dagger)$
\STATE $\rho\gets  \frac{1}{\tr\rho}\rho$
\RETURN $\rho$
\end{algorithmic}
\end{algorithm}

\section{Description of the package}\label{sec:package}
The \TRQS\ package implements a number of functions allowing to obtain random
state vectors and random density matrices. 

It uses Quantis random number generator produced by
ID Quantique~\cite{idquantique} for the purpose of generating random numbers. ID
Quantique provides drivers, example programs and a library for accessing Quantis
devices on most popular operating systems. The software package for using Quantis,
including some examples of how \libquantis\ library can be used, can be
downloaded from the ID Quantique support page~\cite{quantis-support}.

Note that the functions implemented are independent from the actual source of
randomness. In particular it is possible to switch to some other sources of
random numbers.

The package consists of a set of source files, developed using \MathLink\
provided by \Mathematica\ and a package file \texttt{TRQS.m}, implementing the
main functionality.
\subsection{Communication with Quantis device}
In the presented package the communication between \Mathematica\ and
Quantis device was implemented using \MathLink\ -- a standard interface for
interprogramme communication provided by \Mathematica.

We used \libquantis\ library that provides a number of functions for reading random data from Quantis
device. In particular the presented package uses only functions for reading
basic data types. The functions of this kind implemented in \libquantis\ can be
divided into four categories
\begin{itemize}
    \item \texttt{QuantisReadShort} and \texttt{QuantisReadScaledShort} --
    functions for reading short integers and short integers within the given
    range,
    \item \texttt{QuantisReadInt} and \texttt{QuantisReadScaledInt} -- functions
    for reading long integers and long integers within the given range,
    \item \texttt{QuantisReadFloat\_01} and \texttt{QuantisReadScaledFloat}
    functions for reading float numbers in the range $[0,1]$ and float numbers
    within the given range,
    \item \texttt{QuantisReadDouble\_01} and \texttt{QuantisReadScaledDouble} --
    functions for reading double numbers in the range $[0,1]$ and double numbers
    within the given range.
\end{itemize}

Moreover, the library function \texttt{QuantisRead} allows to read raw random
data. This function can be also used to read large amount of data, which can be
necessary in the case when one needs to fill large matrices with random numbers.

\subsection{Organisation of the package}
The described package was designed to work in companion with the QI package for
\Mathematica~\cite{QI} and can be used along with this package. The provided
functions for generating random states can be grouped into three categories:
basic functions, functions for generating pure states and unitary matrices and
functions for generating mixed states and channels. 

The functions are defined within the \TRQS\ name space. We follow the naming
convention which assumes that functions using a true random number generator to
produce results have names starting with \InlineCode{True} and the rest of the
name describes the generated object. The functions related to the configuration
of the back-end \ie\ Quantis device, have names starting with
\InlineCode{Quantis} (see: Sec.~\ref{sec:configuration-functions}).

Each function is provided along with some basic information about its
functionality.

\subsubsection{Basic functions}\label{sec:basic-functions}
The first group of functions implements basic structures utilised for
generating quantum states. The functions in this group implement communication 
with Quantis device and provide the generation of real and integer random numbers 
and some basic structures. In this group the following functions
allow to access basic types of random numbers:
\begin{enumerate}
    \item \texttt{TrueRandomReal} -- returns a random real (double) number; this
    function is based on \libquantis\ library function
    \texttt{QuantumReadScaledDouble} and is implemented in three variants: 
    \begin{enumerate}
        \item \texttt{TrueRandomReal[$\{n_{\min},n_{\max}\}$]} -- returns a real
        number distributed uniformly in the interval $[n_{\min},n_{\max}]$,
        \item \texttt{TrueRandomReal[$n_{\max}$]} -- returns a real number
        distributed uniformly in the interval $[0,n_{\max}]$,
        \item \texttt{TrueRandomReal[]} -- returns a real number
        distributed uniformly in the interval $[0,1]$,
    \end{enumerate}

    \item \texttt{TrueRandomRealNormal[x,y,$\{d_{1},\ldots,d_{l}\}$]} -- returns
    a $(d_{1}\times\ldots\times d_{l})$-dimensional array of random numbers
    distributed according to \NormalDist{x}{y}.
    \item \texttt{TrueRandomInteger} -- returns a random integer. This function,
    based on \libquantis\ library function \texttt{QuantumReadScaledInt}, is
    provided for convenience and implemented in three variants:
    \begin{enumerate}
        \item \texttt{TrueRandomInteger[$\{n_{\min},n_{\max}\}$]} -- returns
        an integer distributed uniformly in the interval $[n_{\min},n_{\max}]$,
        \item \texttt{TrueRandomInteger[$n_{\max}$]} -- returns an integer
        distributed uniformly in the interval $[0,n_{\max}]$,
        \item \texttt{TrueRandomInteger[]} -- returns 0 or 1.
    \end{enumerate}
    
\end{enumerate}
The following functions, built using these basic functions, allow to obtain
the structures used to construct random quantum states and operations:
\begin{enumerate}

    \item \texttt{TrueRandomSimplex[n]} -- returns an element of a standard
    simplex, distributed uniformly on simplex,
    \item \texttt{TrueGinibreMatrix[m,n]} -- returns an $n\times m$ Ginibre
    matrix,
    \item \texttt{TrueRandomChoice[\{$e_1$,$e_2$,\ldots,$e_n$\}]} -- returns 
    at random one of the \{$e_1$,$e_2$,\ldots,$e_n$\}.,
	\item \texttt{TrueRandomGraph[{v,e}, form]} -- returns a pseudorandom graph
	with v vertices and e edges. Additionally, the last argument can be set to
	"Graph" (default) to obtain a graphical representation of the result or to
	"List" to obtain the result as a list of vertices and edges.
\end{enumerate}

\subsubsection{Pure states and unitary matrices}\label{sec:pure-unitary-functions}
The functions in this group allow to obtain random pure states and random unitary
matrices. Since product (or local) states and operations are of a special
interest in quantum information theory, we provide functions allowing to
generate pure states and unitary matrices of the tensor product structure
\begin{enumerate}
    \item \texttt{TrueRandomKet[n]} -- returns a random pure state in
    $n$-dimensional space $\Cplx^n$,
    \item \texttt{TrueRandomProductKet[$\{n_1,n_2,...n_k\}$]} -- returns a
    random pure state, which is an element of space with the tensor product
    structure $\Cplx^{n_1}\otimes\Cplx^{n_2}\otimes\ldots\otimes\Cplx^{n_k}$
    \item \texttt{TrueRandomUnitary[n]} -- returns a random unitary matrix
    acting on $n$-dimensional space $\Cplx^n$,
    \item \texttt{TrueRandomLocalUnitary[$\{n_1,n_2,...n_k\}$]} -- returns a
    random unitary matrix, which acts on the elements of space with the tensor
    product structure $\Cplx^{n_1}\otimes\Cplx^{n_2}
    \otimes\ldots\otimes\Cplx^{n_k}$
\end{enumerate}

\subsubsection{Mixed states}\label{sec:mixed-channels-functions}
The last group of functions implements the generation of random mixed states. In
particular we have:
\begin{enumerate}
    \item \texttt{TrueRandomStateHS[n]} -- a random density matrix of dimension
    $n$, generated according to the Hilbert-Schmidt measure,
    \item \texttt{TrueRandomStateBures[n]} -- a random density matrix of
    dimension $n$, generated according to the Bures measure,
    \item \texttt{TrueRandomStateInduced[n,k]} -- a random density matrix of
    dimension $n$, generated according to the induced probability measure with
    an external system of dimension $k$,
    \item \texttt{TrueRandomProductState[$\{n_1,n_2,...n_k\}$,$\mu$]} -- a
    product random density matrix acting on the space with the tensor product
    structure $\Cplx^{n_1}\otimes\Cplx^{n_2} \otimes\ldots\otimes\Cplx^{n_k}$
    and with each local component generated according to measure $\mu$, where
    $\mu$ can be set to \texttt{"HS"}, \texttt{"Bures"} or some integer $K$
    describing an induced measure.
\end{enumerate}

Additionally \TRQS\ package allows to generate random dynamical matrices, 
representing the most general form of quantum system evolution.
\begin{enumerate}
    \item \texttt{TrueRandomDynamicalMatrix[n,k]} -- a random dynamical matrix
    of dimension $n$, representing a quantum channel acting on
    $n$-dimensional space of density matrices, with $k$ eigenvalues set
    to 0. The last argument is set to 0 by default.
\end{enumerate}
The above function is based on the algorithm described in \cite{bruzda09random}.
The obtained random dynamical matrix can be easily transformed into a set of
random Kraus operators~\cite{BZ06,miszczak11singular}.

\subsubsection{Functions related to the back-end configuration}\label{sec:configuration-functions}
To provide some basic interaction with the underlying device, the following
functions were implemented in \TRQS\ package.
\begin{enumerate}
    \item \texttt{QuantisGetLibVersion[]} -- returns a version number of the
    installed libQuantis library.
    \item \texttt{QuantisGetSerialNumber[]} -- returns a serial number of
    Quantis device used as a back-end.
    \item \texttt{QuantisGetDeviceID[]} -- returns an id number of Quantis
    device.
    \item \texttt{QuantisGetDeviceType[]} -- returns a type of Quantis device.
\end{enumerate}

Note that the functions \texttt{QuantisGetDeviceID[]} and
\texttt{QuantisGetDeviceType[]} provide only information about the configuration
options used during the compilation of \MathLink\ source files.

\section{Examples}\label{sec:examples}
The main aim of the presented package is to provide a tool for the analysis of
the properties of random density matrices. Below we present two examples of such
analysis. First, we calculate the distributions of eigenvalues for 4-dimensional
mixed density matrices and compare analytical and numerical results. Next, we
calculate numerically the average fidelity between random density matrices with
respect to measure $\mu_{2,K}$. In both cases we compare the results obtained
using the presented package and the results obtained from a standard random
number generator with the analytical results.

We also provide a comparison of speed between the standard pseudorandom number
generator from \Mathematica\ and generator using \libquantis\ library. This 
example shows that the speed of random number generation offered by the currently
available hardware is insufficient.

\subsection{Distribution of eigenvalues}
The Bures and Hilbert-Schmidt probability measures are of the product form \ie\
the distribution of eigenvalues is independent from the distribution of
eigenvectors. 

In the case of the Hilbert-Schmidt measure the probability density of
eigenvalues is given by the formula~\cite{BZ06}
\begin{equation}
P_{\mathrm{HS}}(\lambda_1,\ldots,\lambda_N)=
C_N^{\mathrm{HS}} \prod_{i<j}(\lambda_i-\lambda_j)^2,
\end{equation}
where $\sum_i\lambda_i=1$, $\lambda_i\leq0, i=1,2,\ldots,N$. The normalisation
constant $C^{\mathrm{HS}}_N$ reads
\begin{equation}
C_N^{\mathrm{HS}}=\frac{\Gamma(N^2)}{\prod_{i=1}^N\Gamma(k)\Gamma(k+1)}.
\end{equation}

\begin{figure}[ht]
	\centering
	\subfigure[Analytical results obtained by a direct integration over
	appropriate subsets of the convex hull of the spectrum.]{
	\includegraphics[width=0.45\textwidth]{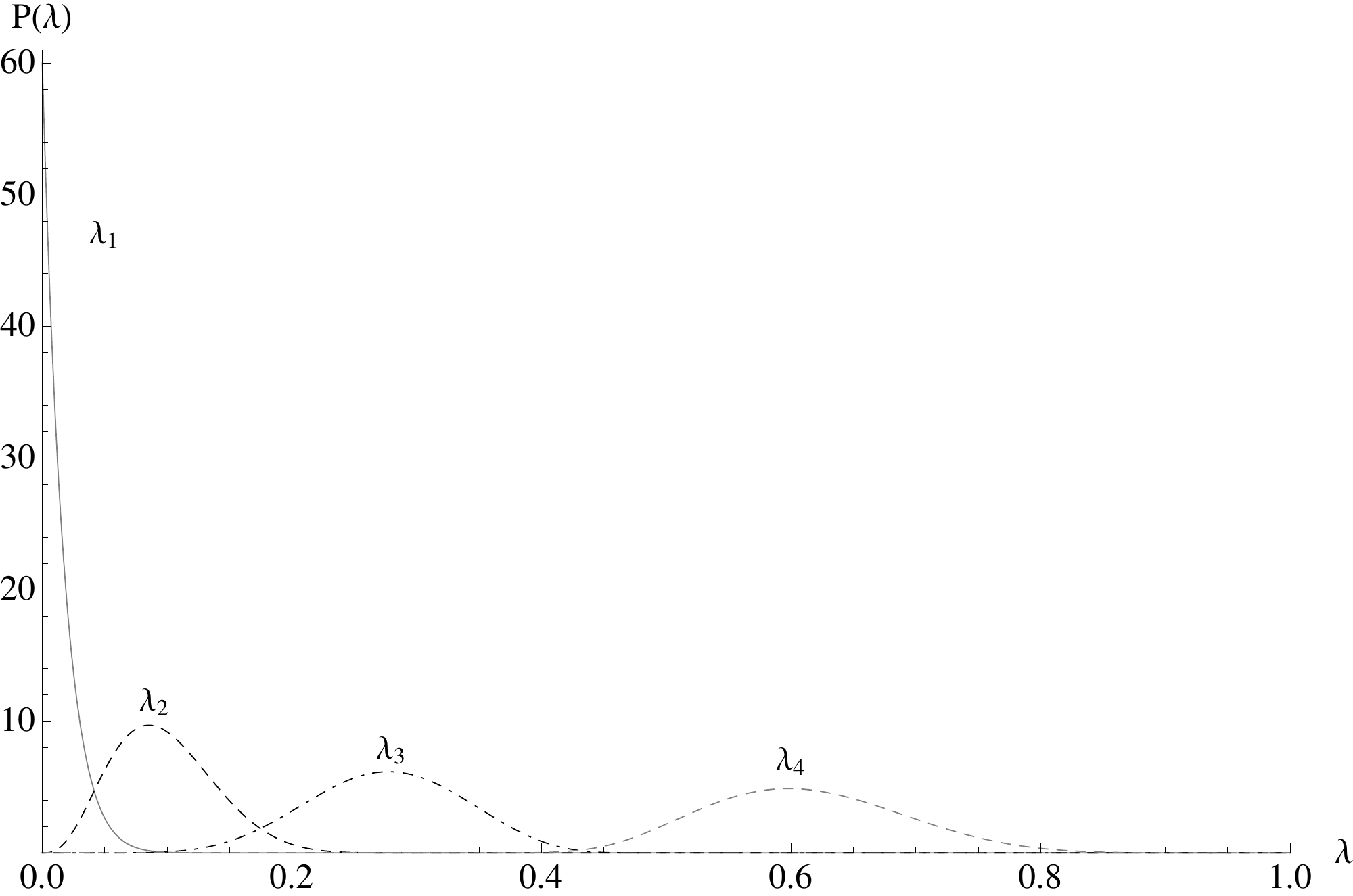}
	\label{fig:lambda-prob-hs}}\quad    
    \subfigure[Numerical results obtained using random states generated with
    TRQS package.]{
    \includegraphics[width=0.45\textwidth]{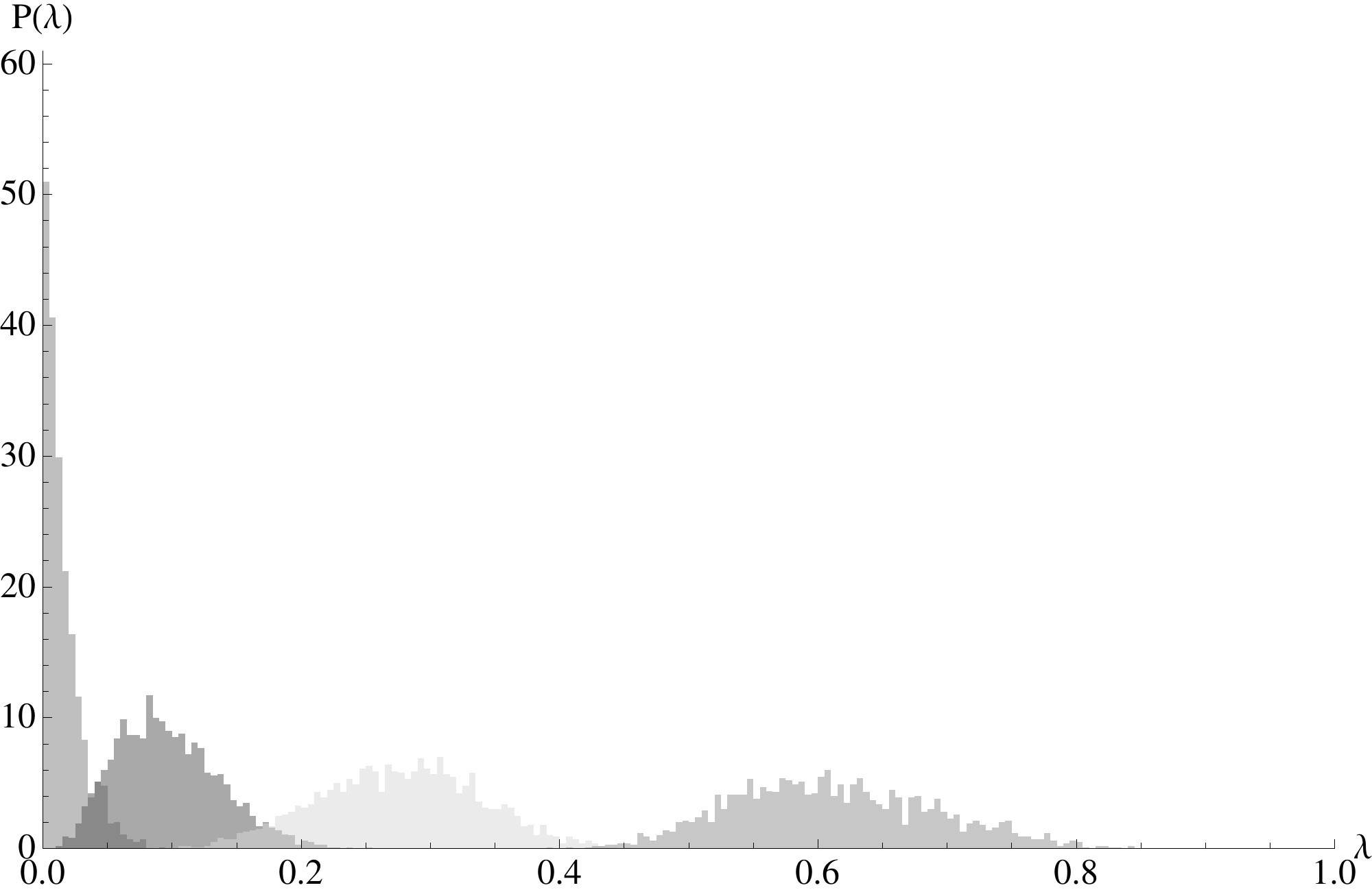}
    \label{fig:lambda-prob-hs-trqs} }
    \caption{Distribution of eigenvalues for random density matrices distributed
    uniformly according to the Hilbert-Schmidt probability measure.}
\end{figure}

Here we present some results for the Hilbert-Schmidt measure and density
matrices of dimension~4. The distribution of eigenvalues
$\lambda_1,\lambda_2,\lambda_3,\lambda_4$ of the random density matrices from
$\Omega_4$ generated uniformly with respect to the Hilbert-Schmidt measure is
presented in Fig.~\ref{fig:lambda-prob-hs}. In
Fig.~\ref{fig:lambda-prob-hs-trqs} the distribution of
$\lambda_1,\lambda_2,\lambda_3,\lambda_4$ obtained using true random density
matrices is presented. Numerical results were obtained using a sample of 2000
random density matrices. 

\subsection{Average fidelity}
Quantum fidelity~\cite{BZ06} is commonly used in quantum information theory to
quantify to what degree a given quantum state can be approximated by some other
state or a family of states~\cite{markham08geometric}.

The average fidelity between two random quantum states can be used \eg\
to provide an insight into the performance of quantum protocols in the presence
of noise. Since, in most cases, in quantum information processing one is
interested in the behaviour of 2-dimensional systems (qubits), below we deal
with this case only.

As it has already been mentioned, the use of random states in quantum
information processing is commonly motivated by the interaction of the system in
question with the environment. In this case one is interested in random density
matrices generated uniformly with respect to some induced measure $\mu_{2,K}$,
where $K$ is the dimension of the ancillary system. 

\begin{figure}[ht!]
    \centering
    \includegraphics[width=0.5\textwidth]{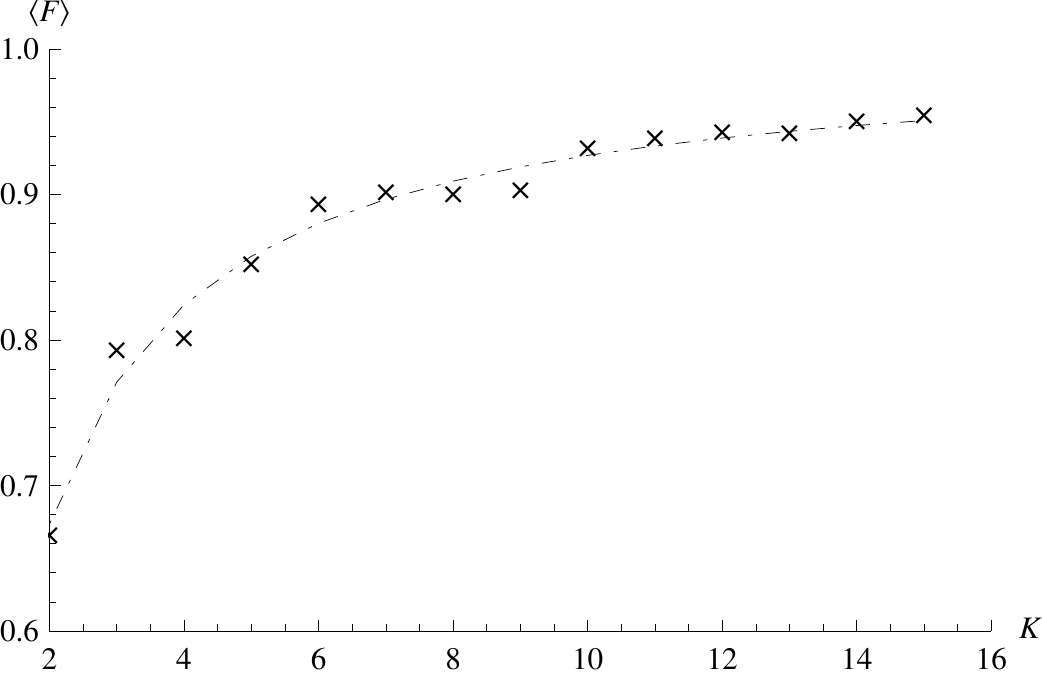}
    \caption{Average fidelity between one-qubit random mixed states generated
    uniformly with respect to $\mu_{2,K}$. The dotted line represents the exact
    result. Numerical results obtained using the presented package are marked
    with "$\times$".}
\label{ref:average-fidelity-induced-k}
\end{figure}

The mean fidelity between two one-qubit random density matrices generated
uniformly with respect to measure $\mu_{N,K}$ was calculated
in~\cite{zyczkowski05averagefidelity} and reads
\begin{equation}\label{eqn:averagefidelity-qubit}
\langle F\rangle_{2,K} = \frac{1}{2} + \frac{1}{2}\left(\frac{\Gamma
\left(K-\frac{1}{2}\right) \Gamma \left(K+\frac{1}{2}\right)}{\Gamma (K-1)
\Gamma(K+1)}\right)^2.
\end{equation}

The average fidelity for one-qubit random states generated with $\mu_{2,K}$ is
presented in Fig.~\ref{ref:average-fidelity-induced-k}. The results were
obtained using a sample of 50 states and one can see that in this case it allows
to obtain a very good approximation of an exact result, especially in the case
of large~$K$.

\subsection{Speed comparison}
For the purpose of testing the speed of the \TRQS\ package we have performed
three experiments involving generation of random real numbers distributed
uniformly on the unit interval. In each experiment we have used a different
method.

\begin{figure}[htp!]
\centering
\includegraphics[scale=0.7]{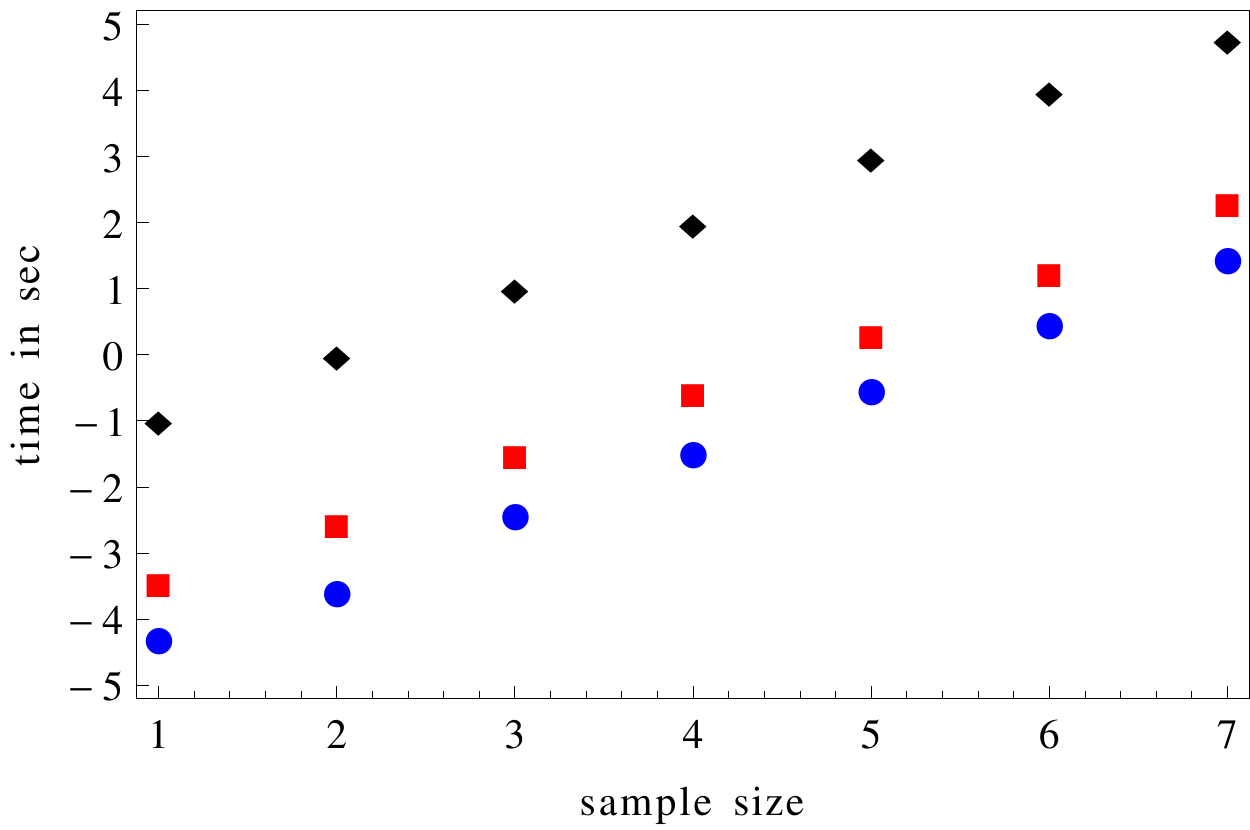}
\caption{Comparison of speed for random number generators in log-log scale. Blue
circles represent timings for samples generated using \texttt{RandomReal[]}
functions using pseudo-random number generator. Red squares represent timings
for samples generated with \TRQS\ package function \texttt{TrueRandomReal[]}
using \MathLink\ executables linked against \libquantisnoWH\ library. Black
diamonds illustrate timings for an analogous method with \MathLink\ executables
linked against \libquantis\ library.}
\label{fig:timingsPlot}
\end{figure}

\begin{enumerate}
	\item The first experiment was conducted using a standard pseudo-random
	number generator provided by \Mathematica. Additionally we used
	\texttt{ClearSystemCache[''Numeric'']} in order to generate the results
	independent from previous computations.
	\item In the second experiment the numbers were generated using \TRQS\
	package and \MathLink\ executables linked against \libquantisnoWH\ library.
	While in this case the generated numbers are still pseudorandom, this test
	was included in order to measure the overhead stemed from the access to an
	external library.
	\item The last experiment was conducted with \TRQS\ package using data
	from the physical quantum random number generator.
\end{enumerate}
In each experiment samples of size $10^1$,$10^2$,\ldots,$10^7$ were generated.

The obtained results are presented in Fig.~\ref{fig:timingsPlot}. The comparison
of timings for samples generated using pseudo-random number generator provided
by \Mathematica\ with timings for data obtained using Quantis device clearly
shows that there is a tremendous difference in the speed of these generators.
For example in order to generate a sample of $10^2$ real numbers using a Quantis
device one needs to wait about 1 sec. Analogous sample is obtained in about
$3\times 10^{-4}$ sec when using a pseudorandom number generator provided by
\Mathematica.

At the same time, the sample of $10^2$ can be obtained using \TRQS\ package if
the used \MathLink\ executables are linked against \libquantisnoWH\ library.
This shows that the main overhead in generating random numbers using Quantis
generator steams from the very slow physical scheme used to obtain random data.

\section{Concluding remarks}\label{sec:conclusions}
Good random number generators are undoubtedly one of the most crucial elements
used in computational physics. In particular, in simulations of quantum
computing the use of random numbers is required to imitate the statistical
behaviour of quantum mechanical objects, \eg~quantum register after
measurement~\cite{oemerMSC2} or particle in quantum
walks~\cite{marquezino08qwalk}. 

The described package can be used along with QI package for
\Mathematica~\cite{QI} and some of the described functions are implemented in QI
with the use of a pseudo-random number generator available in \Mathematica. As
the functions implemented in the presented package operate on basic data types
available in \Mathematica, it is also possible to use the package with other
\Mathematica\ packages developed for the simulation of quantum
computing~\cite{qdensity,quantum2,qcwave}. However, the potential application of
the presented package is not limited to quantum information theory and the
implemented functions can be used in other fields where good quality random
numbers are required.

The obtained timings for different methods of producing random numbers suggest that
the main obstacle in using the presented software in large scale simulations
using random numbers is the speed of the random number generators. Clearly, at
the moment the built-in pseudo-random number generator in \Mathematica\
outperforms the Quantis-based random number generator. Quantis device provides a
stream of random numbers generated at 4 Mbits/s. Additionally, the speed of
random number generation is limited by the speed of the I/O operations. The
speed of functions using Quantis QRNG can be improved by using \libquantis\
function \InlineCode{QuantisRead} for reading a larger amount of random data in
the situation when \eg\ large arrays are filled with random numbers. However,
for the needs of simulations connected to quantum information theory, especially
related to investigations of properties of low-dimensional systems the presented
functions provide a satisfactory user experience. On the other hand, the recent
progress in quantum random generation provides the methods for delivering random
numbers generated at a rate of up to 50 Mbit/s~\cite{furst10high-speed} or
higher~\cite{symul11realtime}.

Clearly the application of the described package is limited by the availability
of Quantis quantum random number generator. However, alternative sources of
random numbers generated using hardware operating on the basis of quantum
mechanics exist. In particular, QRNG Service provided by PicoQuant GmbH and the
Nano-Optics group at the Department of Physics of Humboldt
University~\cite{qrng-de} allows to obtain samples of random numbers generated
using quantum hardware. The samples can be downloaded directly via web page or,
alternatively, using the provided library~\texttt{libQRNG}. This library can be
used in 32 and 64-bit versions of Linux and Windows operating systems. Another
option is provided by the Quantum Random Bit Generator Service \cite{qrbgs-hr}
developed by Centre for Informatics and Computing, Ruder Bo\v{s}kovi\'c
Institute, Zagreb, Croatia. This service provides bindings for a variety of
programming languages, including C, Java and Python. Both services require
registration and have some limitations concerning the amount of random data that
can be downloaded. However, they provide free and relatively easy to use
alternative for the commercial solution provided by ID Quantique. 

\section{Acknowledgements}
Author would like to acknowledge interesting discussions with K.~\.Zyczkowski
and W.~Roga and some stimulating programming exercises conducted with
Z.~Pucha{\l}a and P.~Gawron. Thanks also goes to L.~Widmer for motivating this
work and H.~Weier for his comments concerning the recent progress in the area of
quantum random generation methods. Author would like to thank the anonymous
reviewer for his motivating comments concerning the development of the package.

This work was supported by the Polish National Science Centre under the grant
number N N516 475440 and by the Polish Ministry of Science and Higher Education
under the grants number N N519 442339 and IP 2010 052 270.

\bibliographystyle{unsrt}
\bibliography{quantis_random_states}

\end{document}